\documentclass[%
reprint
,twocolumn%
 ,nobibnotes,pre,floatfix,superscriptaddress]{revtex4-1}
\usepackage{graphicx}
\usepackage{epstopdf}
\usepackage{amsfonts}
\usepackage{amsmath}
\usepackage{bm}
\usepackage{enumerate}
\usepackage{color}
\begin{document}

\title{Delay-induced stochastic bursting in excitable noisy systems}
\author{Chunming Zheng}
\affiliation{Institute for Physics and Astronomy, University of Potsdam, Karl-Liebknecht-Strasse 24/25, 14476 Potsdam-Golm, Germany}
\author{Arkady Pikovsky}
\affiliation{Institute for Physics and Astronomy, University of Potsdam, Karl-Liebknecht-Strasse 24/25, 14476 Potsdam-Golm, Germany}
\affiliation{Research Institute for Supercomputing, Nizhny Novgorod State University, Gagarin Avenue 23, 606950 Nizhny Novgorod, Russia}
\date{\today}

\begin{abstract}
We show that a combined action of noise and delayed feedback
on an excitable theta-neuron leads to rather coherent
stochastic bursting. An idealized point 
process, valid if the characteristic time scales in the problem are well-separated,
is used to describe statistical properties such as the power spectral density and the interspike interval 
distribution.
We show how the main parameters of the point process, the spontaneous excitation rate and the
probability to induce a spike during the delay action, can be calculated from
the solutions of a stationary and a forced Fokker-Planck equation.
\end{abstract}

\maketitle

\section{Introduction}
\label{sec:Introduction}

Time-delayed feedback and noise are factors that substantially 
contribute to complexity of the dynamical behaviors.
While noise generally destroys coherence of oscillations, there are situations
(e.g., stochastic and coherence resonances) where it plays a constructive role leading
to a quite regular behavior~\cite{McDonnel_etal-08,pikovsky1997coherence}. Also delayed feedback can  
either increase or suppress coherence of 
oscillators~\cite{Goldobin-Rosenblum-Pikovsky-03,Goldobin-Rosenblum-Pikovsky-03a,janson2004delayed}.
Interplay of delay and noise is important for neural systems, where it has been
studied both on the level of individual neurons~\cite{Prager_etal-07}, of networks of coupled
neurons~\cite{Kouvaris_etl-10}, and of rate equations~\cite{Goychuk-Goychuk-15}.

A significant progress in understanding of an interplay of noise and delayed feedback
has been achieved for bistable 
systems~\cite{tsimring2001noise,masoller2003distribution}. Furthermore, variants of the 
bistable dynamics with highly asymmetric properties of the two states have been adopted to
describe excitable systems under delay and noise~\cite{Prager_etal-07,Pototsky-Janson-08,Kouvaris_etl-10}. 
In this paper we develop another approach to the dynamics of excitable 
noisy systems with a delayed feedback.
We investigate a theta-neuron model~\cite{ermentrout1986parabolic}, 
which is a paradigmatic 
example of an excitable system in mathematical and computational neuroscience. 
Under the action 
of a small noise, this system 
demonstrates a random, Poisson sequence of 
spikes. For the stochastic excitable theta neuron model, 
the interspike interval distribution and the coefficient of 
variation have been analyzed analytically in 
Refs.~\cite{gutkin1998dynamics,lindner2003analytic}.
We will show that a small additional delayed feedback
(large feedback can significantly modify the dynamics, see, e.g., \cite{kromer2014noise})
leads to an interesting partially coherent spike pattern which we 
call \emph{stochastic bursting}. Bursting describes a general phenomenon with quiescent periods following periods of rapid repeated firing and is thought to be important in communication between neurons and synchronization~\cite{izhikevich2007dynamical}. In our present paper, the bursts themselves
appear at random instants of time and have random duration, 
but inside each burst
the spikes are separated by nearly constant time intervals. Contrary to the
bistable models, in our description we consider only the excitable
state as stochastic one, while the excitation itself is deterministic. 

The paper is organized as follows. We first formulate the basic model
in Section~\ref{sec:Model}. Then, in Section~\ref{sec:point_process} we formulate a point process description
of the stochastic bursting, and derive statistical properties such as the distribution of inter-spike intervals and the
power spectral density. In this description there are two parameters, the rate of excitation and the probability for delayed feedback
to induce a spike. The latter quantity is nontrivial, and we describe approaches to its calculation
in Section~\ref{Sec:probp}. We discuss the results in Section~\ref{sec:concl}.

\section{Model formulation}
\label{sec:Model}

In this paper we study the dynamics of an excitable system 
subject to noisy input and delayed
feedback. 
The model is described by a scalar variable $\theta$ defined on a circle:
\begin{equation}\label{Eq:model}
	\dot{\theta}=a+\cos\theta+\epsilon(a+\cos\theta(t-\tau))+\sqrt{D}\xi(t){\color{red}{.}}
\end{equation}
Here parameter $a$ defines the excitability properties, parameter $D$ describes the level of external noise
(which we assume to be Gaussian white one, $\langle \xi(t)\rangle=0$,   $\langle
\xi(t)\xi(t^{'})\rangle=2\delta(t-t^{'})$), and $\epsilon$ is the amplitude of a delayed feedback.
The feedback is chosen to vanish in the steady state of the system.
Model~\eqref{Eq:model}, without delayed feedback, is very close to the 
theta-neuron model~\cite{ermentrout1986parabolic},
extensively explored in different contexts in neuroscience
(where inclusion of noise is very natural, while a delayed feedback is often
attributed to the autapse effect, cf.~\cite{luke2014macroscopic}), and to the 
active rotator model~\cite{Park-Kim-96,*Tessone_etal-07,*Zaks_etal-03,*Sonnenschein-13,*ionita2014physical}.
In~\eqref{Eq:model} we assume a simple additive action of the delayed feedback and of noise. For
theta-neurons, one quite often explores multiplicative forcing, where the force terms are multiplied
with factor $(1-\cos\theta)$ (cf.~\cite{Borgers-Koppell-05},
notice that our variable is shift by $\pi$ to the variable used in \cite{Borgers-Koppell-05}). However, as will be clear from the analysis below,
this brings only small quantitative corrections to the results, while the main qualitative conclusions
remain valid -- because the most sensitive to forcing region in the phase space is around $\theta\approx -\pi$,
and in this domain the factor $(1-\cos\theta)$ is nearly a constant.

For $|a|\lesssim 1$ the autonomous theta-neuron (without noise and feedback) is in an excitable regime:
there are two nearby stationary states, one stable and one unstable.  Both noise and the feedback can kick
the system from the stable equilibrium so that it produces a ''spike''. Our goal in this paper is to 
describe statistical properties of the appearing spike train. Prior to the full analysis, we
briefly outline relatively simple cases of the 
purely deterministic dynamics (no noise) and of the purely noisy dynamics
(no delayed feedback).

\subsection{Deterministic case}
\begin{figure}
	\centering
	\includegraphics[width=\columnwidth]{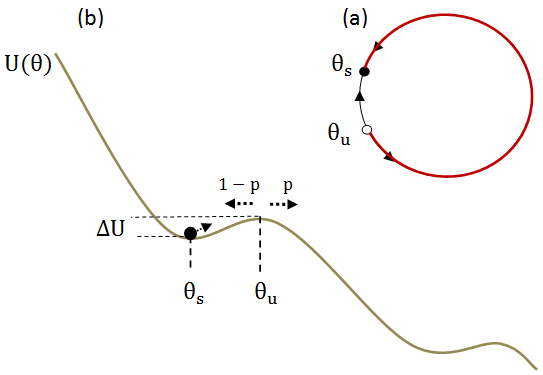}
	\caption{A sketch of the theta-neuron model Eq.~\eqref{Eq:model}. 
	In (a), the red trajectory from $\theta_{u}$ to $\theta_{s}$ 
	represents a spike, while the black curve shows relaxation without a spike. 
	Panel (b) depicts how 
	the 'phase particle' evolves in the effective potential  $U(\theta)$, either
	overcoming the barrier (with probability $p$), 
	or returning back to the equilibrium $\theta_{s}$ (with probability $1-p$).}
	\label{fig:schem}
\end{figure}

\begin{figure}
	\centering
	\includegraphics[width=\columnwidth]{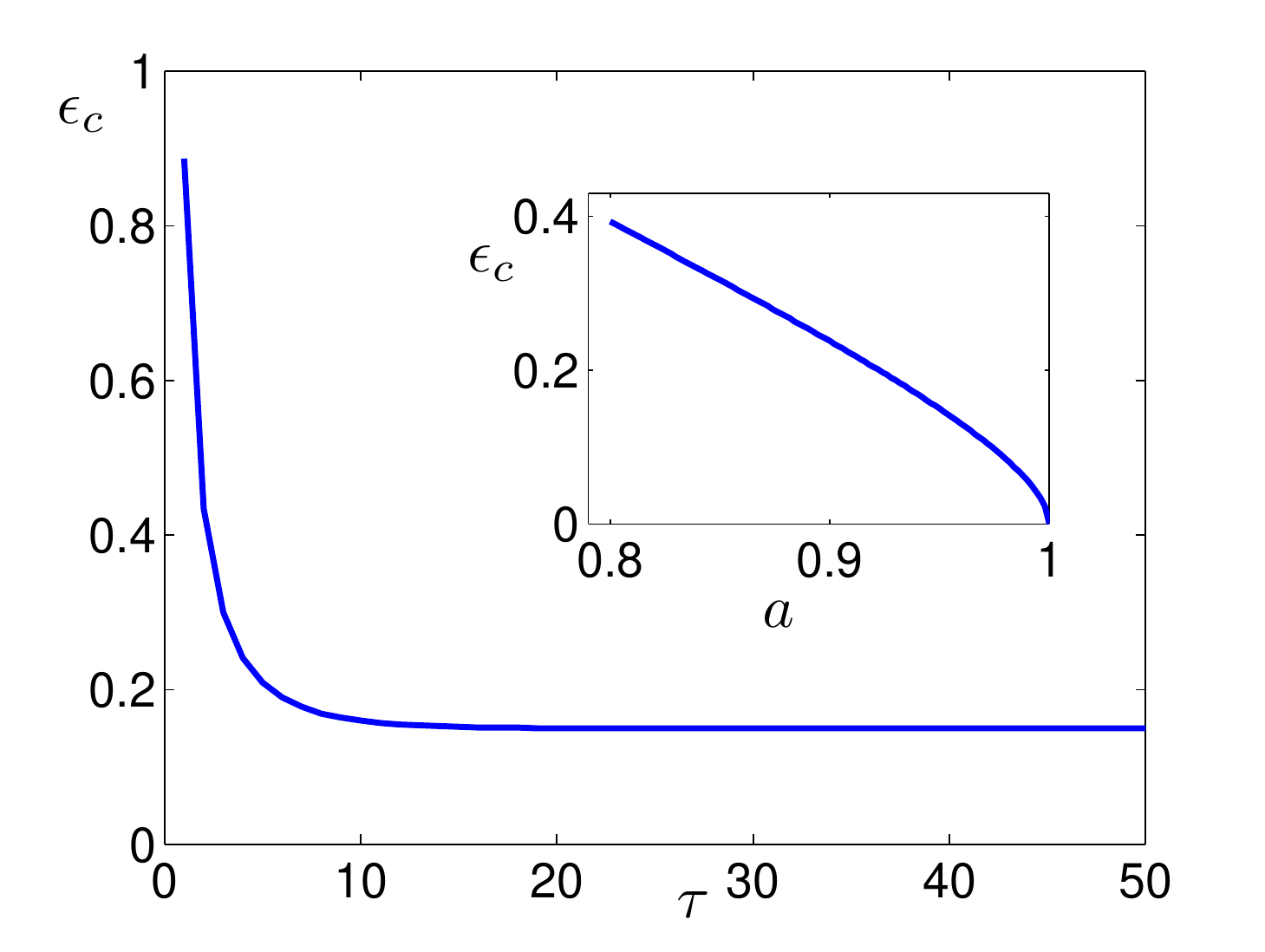}
	\caption{Critical value of $\epsilon$ in dependence on the delay time $\tau$ 
	in the deterministic case with $a=0.95$. The inset is the  
	asymtotic value at large delay times $\epsilon_{c}$ in dependence on
	excitability parameter $a$.}
	\label{Fig:epscr}
\end{figure}

An autonomous theta-neuron (one sets  $\epsilon=D=0$ in~\eqref{Eq:model}) with $|a|\lesssim 1$ is
an excitable system with one stable fixed point at $\theta_{s}=\arccos(-a)$ and another 
unstable fixed point at $\theta_{u}=2\pi-\arccos(-a)$.  One can represent the dynamics as an overdamped motion
in an inclined periodic potential
\begin{equation}\label{Eq:pot}
	\dot{\theta}=-\frac{dU}{d\theta},\qquad U(\theta)=-a\theta-\sin\theta\;,
\end{equation}
for which $\theta_s$ is a local minimum and $\theta_u$ is a local maximum, see Fig.~\ref{fig:schem}.
As parameter $a$ 
is close to the value of a SNIC bifurcation $a=1$, the distance $\theta_{u}-\theta_s$ is small
(correspondingly, the barrier of the potential is small as well)
and already a small external perturbation can produce a nearly $2\pi$-rotation of $\theta$. The 
form of the spike can be represented as a trajectory that starts at $\theta_u$, ends
at $\theta_s$, and reaches the maximal value at time instant $t_0$:
\begin{equation}
\Theta_{sp}(t)=2\arctan \left(\sqrt{\frac{1+a}{1-a}}\tanh\left(\frac{\sqrt{1-a^2}}{2}(t-t_{0})\right)\right)\;.
\label{Eq:spike}
\end{equation}
Let us now consider deterministic model~\eqref{Eq:model} with delay, 
i.e., the case $D=0$. The system still has
a locally stable equilibrium $\theta_s$. However, for large enough $\epsilon$ it can possess stable periodic
oscillations. Indeed, a perturbation of the equilibrium can result in a spike \eqref{Eq:spike}. After the delay
time $\tau$, a force   
\begin{equation}
\epsilon H(t)=\epsilon(a+\cos\Theta_{sp}(t))
\label{Eq:forc}
\end{equation}
will act on the theta-neuron. For a sufficiently  large value of $\epsilon$ it will produce a new spike, etc.
In Fig.~\ref{Fig:epscr}
we show critical values of $\epsilon$ that depend on the delay time $\tau$ as well as the excitability
parameter $a$. Clearly, $\epsilon_c\to 0$ if the excitability parameter $a$ approaches the bifurcation
value $a_{SNIC}=1$. Dependence on the delay time is also rather obvious: for large delays the critical
value $\epsilon_c$ is delay-independent, while for delays comparable to the pulse duration
(which is, according to \eqref{Eq:spike}, $\sim(1-a^2)^{-1/2}$)  there is a blocking effect which mimics
a refractory period for a neuron after a spike.

\subsection{Noisy case}
\label{sec:nc}
If there is no time-delay feedback, i.e., $\epsilon=0$, but noise is present, $D>0$,
the spikes can be induced by noise. The model is well-described in the literature~\cite{risken1996fokker},
here we briefly outline the  features required for
consideration of the more complex case with delay. The dynamics is especially simple for small noise: 
in this case,
most of the time the system stays in a neighbourhood of the stable state $\theta_s$, and the 
excitations are rare. The sequence of spikes builds  a Poisson process with a constant 
spiking rate $\lambda$,  which is equal to the probability current $J$ of the corresponding Fokker-Planck equation
\begin{equation}
\begin{aligned}
\frac{\partial P(\theta,t)}{\partial t}=&-\frac{\partial J}{\partial\theta}=\\
&-\frac{\partial }{\partial\theta}\left[(a+\cos\theta)
P(\theta,t)\right]+D\frac{\partial^{2} P(\theta,t)}{\partial\theta^{2}}.
\end{aligned}
\label{Eq:fpe}
\end{equation}
The stationary solution of \eqref{Eq:fpe} is
\begin{equation}
P_{st}(\theta)=C\int_{\theta}^{\theta+2\pi}\frac{d\psi}{D}e^{-\int_{\theta}^{\psi}\frac{a+\cos\varphi}{D}d\varphi}.
\label{Eq:stationary_withoutdelay}
\end{equation}
Here $C$ is the normalization constant, so the current is represented as
\begin{equation}
\lambda=J=C\left(1-e^{-\int_{0}^{2\pi}\frac{a+\cos\theta}{D}d\theta}\right). \label{Eq:current}
\end{equation}
In the limit of small noise, this exact expression reduces to the Kramers escape rate
over the potential barrier: $\lambda\approx \frac{1}{2\pi}\sqrt{U^{''}(\theta_{s})|U^{''}(\theta_{u})|}\exp\{-\Delta U/D\}$.

\section{Delay and noise induced bursting as a point process}
\label{sec:point_process}

\begin{figure}
	\centering
	\includegraphics[width=\columnwidth]{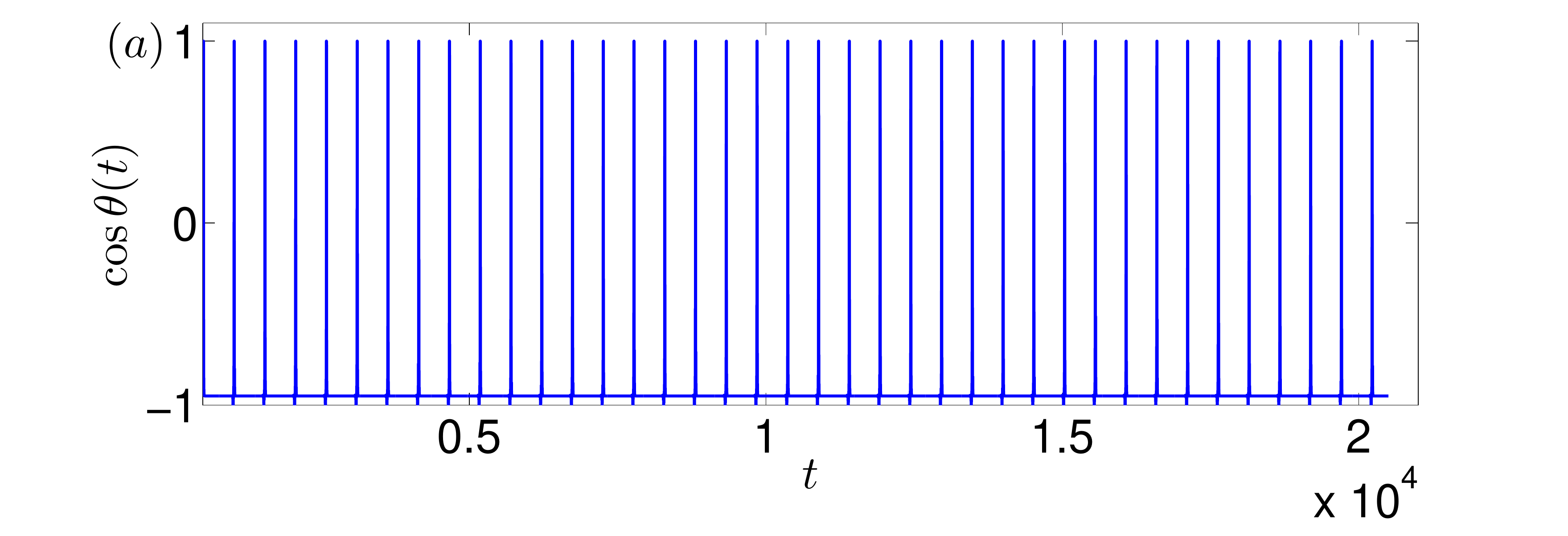}
	\includegraphics[width=\columnwidth]{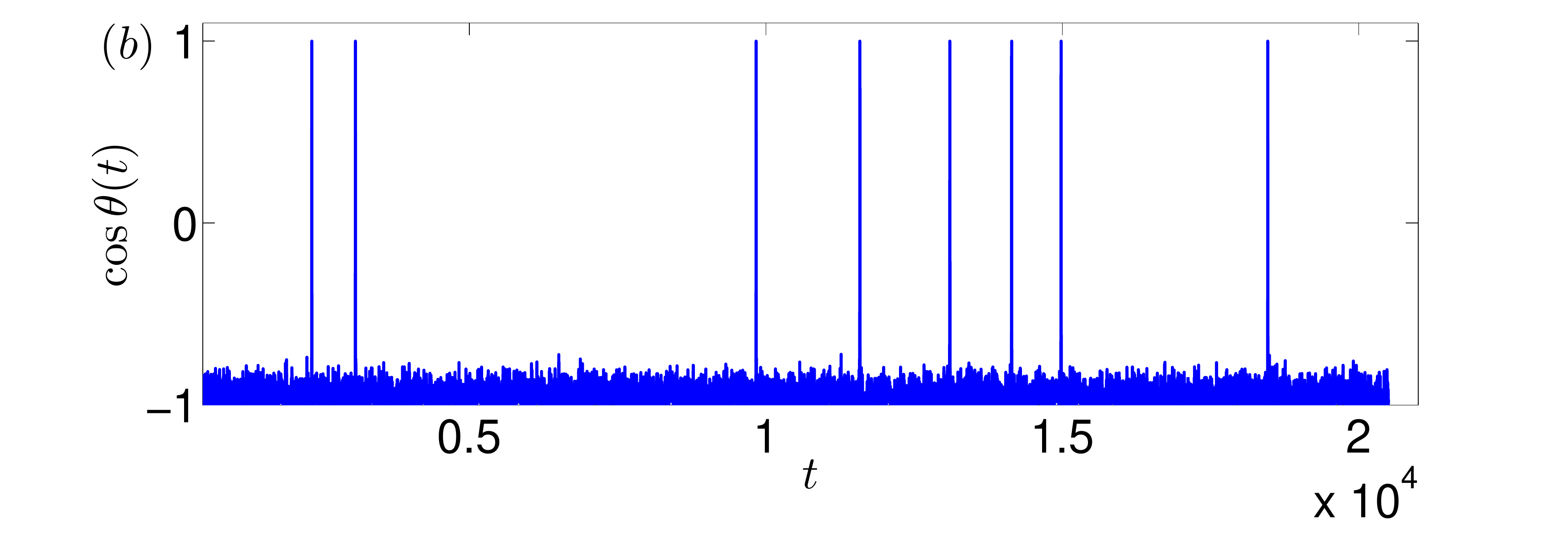}
	\includegraphics[width=\columnwidth]{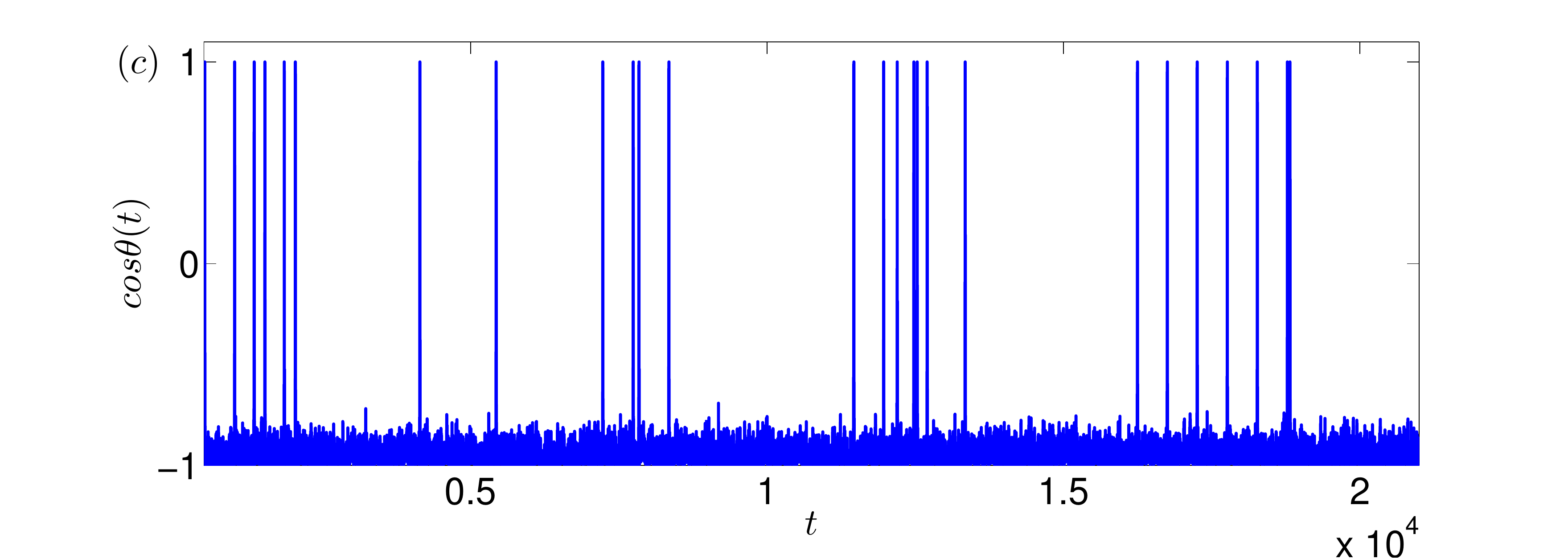}
	\caption{Panel (a) is the periodic solution in the deterministic case when $\epsilon>\epsilon_{c}$. In this deterministic case we simulate Eq.~(\ref{Eq:model}) by the Euler method with discret time interval $dt=0.01$.
	Panel (b) shows  the noisy case without delayed feedback, i.e $\epsilon=0$, with the spike train obeying 
	the Poisson statistics. Panel (c) is the case with both noise ($D=0.005$) and delayed
	feedback ($\epsilon=0.14$) , showing the $\emph{stochastic bursting}$ phenomenon. Spike trains in panel (b) and (c) are obtained through simulation Eq.~(\ref{Eq:model}) by the Euler-Maruyama method with discrete time interval $dt=0.01$. 
	Many spikes with the interval close to $\tau$ are induced, depending on the delay 
	force amplitude. The parameters are $a=0.95, \tau=500$. }
	\label{fig:spike}
\end{figure}

Our main interest here is in the combination effect of time delay and noise with $D\neq 0, \epsilon\neq 0$.
We illustrate the dynamics in Fig.~\ref{fig:spike}(c), where we compare it with the purely
periodic dynamics in the deterministic case (panel (a)) and with the Poisson sequence of spikes
for delay-free case (panel (b)). In panel (c) one can see randomly appearing spikes, like in case (b), and ``bursts''
of several spikes separated by the delay time $\tau$ (like in case (a)). 
Qualitatively, this picture illustrates the two
sources of spike formation: (i) due to a fluctuation of the noise driving, this source is delay-independent,
and (ii) delay-induced spikes which appear due to a combinational effect of delay forcing and noise.
We call the former spikes `spontaneous' ones, or `leaders', and the latter spikes as `induced' ones, or 'followers'.
An exact analytic approach to the noisy dynamics is hardly possible, because 
in presence of delay feedback and noise, the system is non-Markovian.   
Therefore we will next formulate an idealized point process model, which generalizes
the Poisson point process in absence of the delayed feedback. Then, in Section~\ref{Sec:probp} 
we will discuss how to calculate parameters of this point process.
Since the possibility of applying the point process model is based on the separation of time
scales, it is required that the length of the pulse is much smaller than the characteristic
inter-spike interval, which is either the delay time, or the characteristic time interval
between the spontaneous spikes. We assume this conditions to be fulfilled, and use in numerical
examples the parameters that ensure the time scale separation.

\subsection{Point process model}
\label{sec:pp}
Point processes are widely used to mathematically model physical processes that 
can be represented as a
stochastic set of events in time or space, including spike trains. The spike 
train can be viewed as a sequence of pulses, fully determined via
the spike appearance times $t_{j}$. In the case each spike is considered
as a $\delta$-pulse, we have
$\sum_{j}\delta(t-t_{j})$; more generally we can write 
$\sum_{j} H(t-t_{j})$, where $H$ is the waveform~\eqref{Eq:forc}.
In our model, we adopt the leader-follower relationship to describe 
the spiking pattern of type shown in Fig.\ref{fig:spike} (c). 
The spikes which appear when the delay feedback is weak, i.e. solely 
due to a large fluctuation of noise, we call ``spontaneous'' 
ones. As delay plays no role for these spikes, they form a Poisson process with rate 
$\lambda$, as described in Sec.~\ref{sec:nc}.
Each spontaneous spike produces, after delay time $\tau$, forcing \eqref{Eq:forc}. 
During this pulse forcing,
the potential barrier decreases and there is an additional enlarged 
probability to overcome the barrier and to
produce a ``follower'' spike. We denote the total probability to induce the follower spike as $p$
(correspondingly, the probability to have no follower is $1-p$). Of course, each induced 
spike can also produce a follower, with the same probability $p$.  Thus, a leader spike
induces a sequence of exactly $L$ followers with probability $\varrho(L)=p^{L}(1-p)$.

\begin{figure}
	\includegraphics[width=\columnwidth]{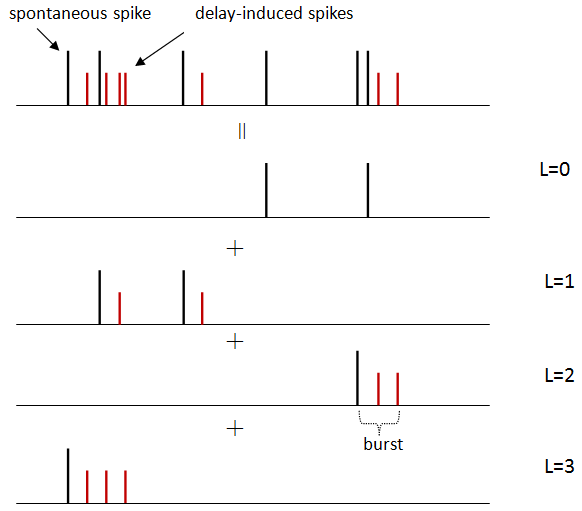}
	\caption{Schematic description of the point process. The black high pulses
	represent the spontaneous spikes (leaders) while the red low ones represent 
	the delay-induced spikes (followers) (the difference in the height of spikes is just 
	a schematic way to classify the events into leaders and followers, while they 
	are of the same height in reality). A leader with a random number of its 
	followers form a burst. The whole process can be viewed as a superposition of sub-processes
	with a fixed number of followers.}
	\label{fig:pointprocess}
\end{figure}

The two parameters, $\lambda$ and $p$, fully describe the point process, consisting of ``bursts''
as shown in Fig.~\ref{fig:pointprocess}. Each burst starts with a leader, which appears with a constant rate
$\lambda$, these leaders form a Poisson process. The followers are separated by the time interval $\tau$, their
number in the burst is random according to the 
distribution   $\varrho(L)$. Noteworthy, the bursts can overlap.

Below we discuss
statistical properties of the point process following from the described model.
It is rather simple to obtain the overall density of spikes.
 Indeed, the average number of followers of a 
 leader is $\sum\limits_{L=0}^{\infty}L\varrho(L)=\frac{p}{1-p}$, and hence the overall spike 
 rate is 
 \begin{equation}
 \mu=\lambda(1+\frac{p}{1-p})=\frac{\lambda}{1-p}\;.
 \label{Eq:mu}
 \end{equation}
Because the process is stationary, the probability to have a spike in a small time interval $(t,t+\Delta)$
does not depend on $t$ and is equal to $\mu \Delta$. Correspondingly, the probability that in
a finite time interval $T$ there is no one spike is $\exp[-\mu T]$.

 \subsection{Interspike interval distribution}
\begin{figure}
	\centering
	\includegraphics[width=0.5\textwidth]{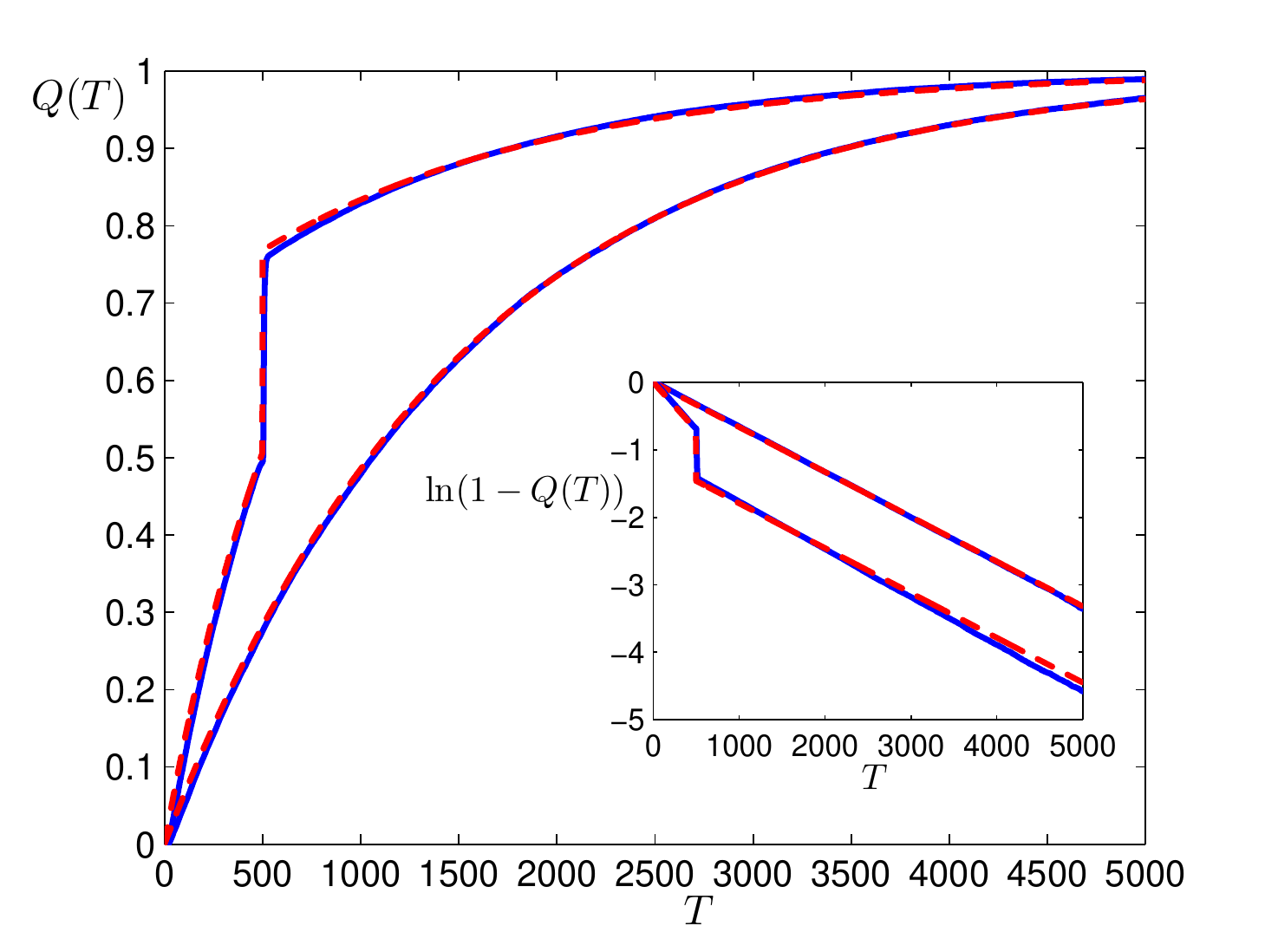}
	\caption{Cumulative ISI distribution $Q(T)$ vs $T$. The blue curve 
	shows numerical simulations of Eq.~(\ref{Eq:model}),  the dashed red curve corresponds to 
	the point process with Eq.~(\ref{Eq:cumulative}), where $\lambda=6.64\times10^{-4}$ is calculated
	from Eq.~(\ref{Eq:current}) and $p=0.53$ is calculated from Eq.~(\ref{Eq:p_numerical}).
	The upper two curves with a jump at $T=\tau$ correspond to the delay case 
	with $\epsilon=0.14$, while the  lower two ones correspond to the case 
	without delay, i.e $\epsilon=0$. The inset in a logarithmic scale
	is to show the coincidence of the 
	slopes, which validates the point process representation of the original model. 
	Parameters are $a=0.95, D=0.005$ and $\tau=500$. }
	\label{fig:ISI}
\end{figure}

Now we derive the interspike interval (ISI) distribution, 
employing the renewal theory \cite{cox1967renewal,gerstner2014neuronal}. Given a spike 
at time $t$ and the next spike at time $t'$, the probability to have no spike 
in the interval $[t, t']$ is called survival function. 
Let us separate the ISI, i.e., $T=t'-t$, into three different cases, 
namely, $T>\tau, T=\tau$ and $T<\tau$. If $T<\tau$, the spikes at $t$ and $t'$ 
can be either spontaneous (leader) or delay-induced ones (followers of spikes preceding that at
$t$), so the survival function is
determined by the full rate $\mu$:  $S(T)=\exp(-\mu T)$. 
In contradistinction, for the case $T>\tau$, the next spike can be only a spontaneous one.
The probability that there is no spike in $[t, t']$ is  the product of three terms: 
the probability to have  no 
spikes in the interval $[t, t+\tau)$ with survival function $S_{\tau b}=\exp(-\mu \tau)$, the probability
$(1-p)$ not to have a follower for the spike at $t$, and the probability
to have no spike in the interval $[t+\tau, t']$, where only the spontaneous 
rate $\lambda$ applies with the survival function  $S_{\tau a}=\exp(-\lambda(T-\tau))$. 
Thus, the survival function for the case $T>\tau$ is $S(T)=S_{\tau b}(1-p)S_{\tau a}=
(1-p)e^{-\mu\tau-\lambda(T-\tau)}$.
Based on the above description and the relationship between the cumulative 
ISI distribution $Q(T)$ and the survival function  $Q(T)=1-S(T)$, the cumulative 
ISI distribution can be obtained as follows:
\begin{equation}  
Q(T)=\begin{cases}
1-e^{-\mu T}, & T<\tau, \\  
1-(1-p)e^{-\mu\tau-\lambda(T-\tau)}, & T\geq \tau.    
\end{cases}       \label{Eq:cumulative}
\end{equation} 
According to the relationship between the cumulative ISI distribution and the ISI 
distribution density $P(T)=Q'(T)$, we can also obtain the ISI distribution density:
\begin{equation}  
P(T)=\left\{  
\begin{array}{lr}  
\mu e^{-\mu T}, & T<\tau, \\  
pe^{-\mu\tau}\delta(T-\tau), & T=\tau,\\  
\lambda(1-p)e^{-\mu\tau-\lambda(T-\tau)}, & T>\tau.    
\end{array}  
\right.  
\end{equation}
We compare the obtained ISI distribution with the numerical result in Fig.~\ref{fig:ISI}.

\subsection{Power spectral density}
Next, we discuss correlation properties of the point process. 
The spike train in our model can be represented as  a superposition of sub-trains having a fixed
number $L$ of followers, see Fig.~\ref{fig:pointprocess} for an illustration of this superposition.
Let us denote $H(t)$ the shape of a spike (it is a delta-function for the point process model, but
for a real process it is given by~\eqref{Eq:spike}).
Then the time series can be written as sum of sub-series of bursts of size $L+1$:
\begin{equation}
x(t)=\sum\limits_{L=0}^{\infty}G_{L}(t)\otimes Y_{L}(t)\otimes H(t)
\end{equation}
where terms $G_L$ and $Y_L$ describe the leaders and the followers for the bursts of size $L+1$:
\begin{equation}
G_{L}(t)=\sum\limits_{i}^{}\delta(t-t_{L}^{i});\quad
Y_{L}(t)=\sum\limits_{l=0}^{L}\delta(t-l\tau)\;.
\end{equation}
The leaders of a sub-series of bursts of size $L+1$
form a Poisson process with the 
rate $\lambda \varrho(L)$,  and the followers 
form a periodic set of spikes with 
separation $\tau$. Here symbol $\otimes$ denotes a convolution.

According to the property of convolution and the independence of the 
sub-series for different $L$, the power spectral density is the sum of spectral densities of the series;
inside each sub-series we have a product of spectral functions:
\begin{equation}
S_{x}(\omega)=\sum\limits_{L=0}^{\infty}S_{G_{L}}
(\omega)S_{Y_{L}}(\omega)S_{H}(\omega)\;.
\end{equation}
Here $S_{G_{L}}(\omega)$ is the power spectral density of the spontaneous spikes, which have the Poisson statistics. 
The power spectral density of the Poisson process is a constant~\cite{stratonovich1967topics}:
\begin{equation}
S_{G_{L}}(\omega)=\lambda \varrho(L)=\lambda (1-p)p^{L}\;.
\end{equation}
The term $S_{Y_{L}}(\omega)$ is the power spectral density of the set of $L$ points separated by
time interval $\tau$, i.e
\begin{equation}
S_{Y_{L}}(\omega)=\left|\int_{0}^{\infty}Y_{L}(t)e^{-i\omega t}dt
\right|^2=\frac{1-\cos(L+1)\omega\tau}{1-\cos\omega\tau}.
\end{equation}
Finally, $S_{H}(\omega)$ is the power spectral density of the shape function 
\[
S_H(\omega)=\left|\int_{-\infty}^{\infty}H(t)e^{-i\omega t}dt\right|^2\;.
\]
Summarizing, we obtain the following expression for the power spectral density of the spike train
\begin{equation}
\begin{aligned}
S_{x}(\omega)&=\sum\limits_{L=0}^{\infty}\frac{1-\cos(L+1)\omega \tau}{1-\cos\omega\tau}\lambda (1-p)p^{L}S_{H}(\omega)\\
&=\frac{\lambda (1+p)}{1+p^2-2p\cos \omega \tau}S_{H}(\omega). \label{Eq:powerspectrum}
\end{aligned}
\end{equation}

The most important part of the spectrum is the first factor, thus we discuss the spectrum for the case
of $\delta$-pulses $S_H=1$.
For the limiting delay-free case, when $p=0$, we have 
$S_{x}(\omega)=\lambda S_H(\omega)$, which corresponds 
to a purely Poisson process of spontaneous spikes. In another limiting case of 
extensive bursting $p\to 1$, the power spectral density becomes a 
periodic sequence of narrow Lorentzian-like peaks at frequencies $\omega=0,\frac{2\pi}{\tau},\frac{4\pi}{\tau},\ldots$.
The width of a peak is $\sim (1-p)$, while the amplitude scales $\sim (1-p)^{-2}$ (the total power of a peak diverges in this
limit because the density of spike diverges).

\begin{figure}
	\centering
	\includegraphics[width=0.5\textwidth]{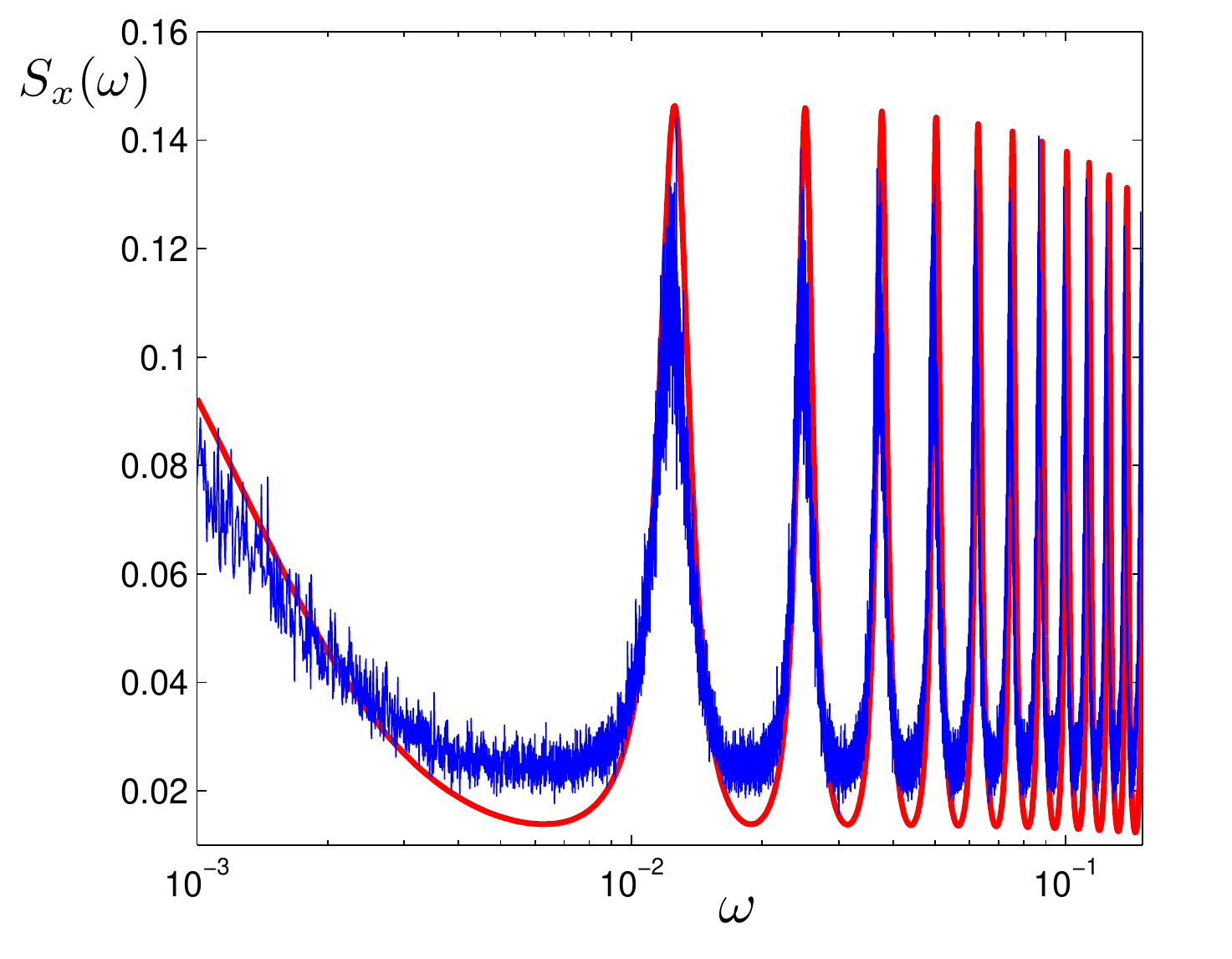}
	\caption{The power spectral density from the simulations (blue curve) and from the point process (red curve)
	described by Eq.(\ref{Eq:powerspectrum}), in which $\lambda=6.64\times10^{-4}, p=0.53$ are 
	calculated from Eq.(\ref{Eq:current}) and Eq.(\ref{Eq:p_numerical}) respectively. The 
	shape function is as Eq.~(\ref{Eq:forc}) describes. Values of $a, D$ and $\tau$ are the 
	same as in Fig.~(\ref{fig:ISI}), i.e $a=0.95, D=0.005$ and $\tau=500$. }
	\label{fig:ISI_power}
\end{figure}

In Fig.~\ref{fig:ISI_power} we compare the obtained expression for the spectral density with direct
numerical modeling of Eq.~\eqref{Eq:model}.

\section{Probability to induce a spike}
\label{Sec:probp}
As have been shown in the section~\ref{sec:point_process} above,
in our model, from the viewpoint of a point process, there are only two parameters:
the spontaneous spiking rate $\lambda$ (or $J$) and $p$, the probability to induce a spike by a 
delay force and noise. The expression for $\lambda$ is given by formula \eqref{Eq:current}. The main challenge
that is discussed in this Section, is an analytical calculation of $p$. 

From the simulations of Eq.~(\ref{Eq:model}), where the delay 
force can be switched off and on 
(corresponding to $\epsilon=0$ and $\epsilon\neq 0$ respectively), 
the probability to induce a spike follows from the relation  \eqref{Eq:mu}:
\begin{equation}\label{Eq:p_simulation}
p=\frac{\langle n\rangle -\langle n_{0}\rangle }{\langle n\rangle }\;.
\end{equation}
 Here $\langle n_{0}\rangle $ is the average number of spikes 
within a large time interval without  the time-delayed force, while $\langle n\rangle$ 
is the average number of spikes in presence of the delayed force within the same time interval. 

\subsection{Induced probability from the solution of the Fokker-Planck equation}

Due to the nolinear force and non-Markovian property of Eq.~\eqref{Eq:model}, 
it's hard to obtain the exact solution analytically, e.g., formulating it in terms 
of delay Fokker Planck equation.
However, since $a$ is close to 1 and the noise intensity is small, we can approximate the 
delay force with a deterministic time-dependent force based on the spike 
solution~\eqref{Eq:spike},\eqref{Eq:forc}. 
Thus, the problem reduces to consideration of a deterministically driven  stochastic model
\begin{equation}\label{Eq:model2}
	\dot{\theta}=a+\cos\theta+\epsilon H(t)+\sqrt{D}\xi(t){\color{red}{.}}
\end{equation}
where the force term is given by expression~\eqref{Eq:forc}. The corresponding
Fokker-Planck equation reads 
\begin{equation}
\frac{\partial P(\theta,t)}{\partial t}=-\frac{\partial }{\partial\theta}\left[(a+\cos\theta+\epsilon H(t)) 
P(\theta,t)\right]+D\frac{\partial^2 P(\theta,t)}{\partial\theta^2}\;.
\label{Eq:nstfp}
\end{equation}

In order to properly formulate the setup for this equation, we need to describe its dynamics qualitatively.
As a starting state prior to incoming pulse $H(t)$, we can take a stationary distribution of the
equation with $\epsilon=0$, i.e. the stationary solution  \eqref{Eq:stationary_withoutdelay}:
$P(\theta,-T)=P_{st}(\theta)$, for $0\leq\theta<2\pi$. Here $-T$ is a starting point of pulse action.
Under action of the pulse, this state evolves, and $P(\theta,t)$ shifts in positive direction of $\theta$,
and a flux of probability through the point $\theta=0$ increases -- this exactly describes
increased local rates of a spike excitation during the action of the pulse. In order to control
``multiple'' pulse excitation (generation of two or more spikes during one acting pulse)
it is convenient to choose the period of domain as $8\pi$ instead of $2\pi$.  
Then, after the action of the pulse $H(t)$, a state $P(\theta,T)$ is reached. 
The net 
probability within the domain $[2\pi, 4\pi]$ can be interpreted as the 
probability to induce just one spike by the force $\epsilon H(t)$ as follows,
\begin{equation}\label{Eq:p_numerical}
p=\int_{2\pi}^{4\pi}(P(\theta,T)-P_{0}(\theta,T))d\theta.
\end{equation}
Here $P(\theta,T)$ is the solution of Eq.~\eqref{Eq:nstfp},
while $P_{0}(\theta,T)$ is the corresponding solution of the unforced  
Fokker-Planck equation (i.e., of Eq.~\eqref{Eq:nstfp} with $\epsilon=0$) -- it describes spontaneous spikes.
The total probabilities in domains 
$[4\pi, 6\pi]$ and $[6\pi, 8\pi]$ (they correspond to the probabilities to induce 2 or 3 spikes) are 
actually very close to zero and therefore can be neglected.

Practically, we solve Eq.~\eqref{Eq:nstfp} with a spectral method. We represent the probability density  as a (truncated) Fourier series as $P(\theta,t)=\sum_{m=-N}^{N}C_{m}(t)e^{i\frac{m}{4}\theta}$, and substitute 
 it into the Fokker-Planck equation. In this way 
 we obtain an large system
of non-autonomous ODEs for the Fourier modes
\begin{equation}
\frac{dC_{m}}{dt}=\frac{m}{8i}C_{m-4}-(\frac{i}{4}ma+\frac{i}{4}m\epsilon H(t)+
\frac{m^{2}}{16}D)C_{m}+\frac{m}{8i}C_{m+4}.
\end{equation}
We truncated this system at $N=400$ and solved the 
above ODEs by the 4th order Runge-Kutta method with  time step $0.001$.

\begin{figure}
	\includegraphics[width=0.5\textwidth]{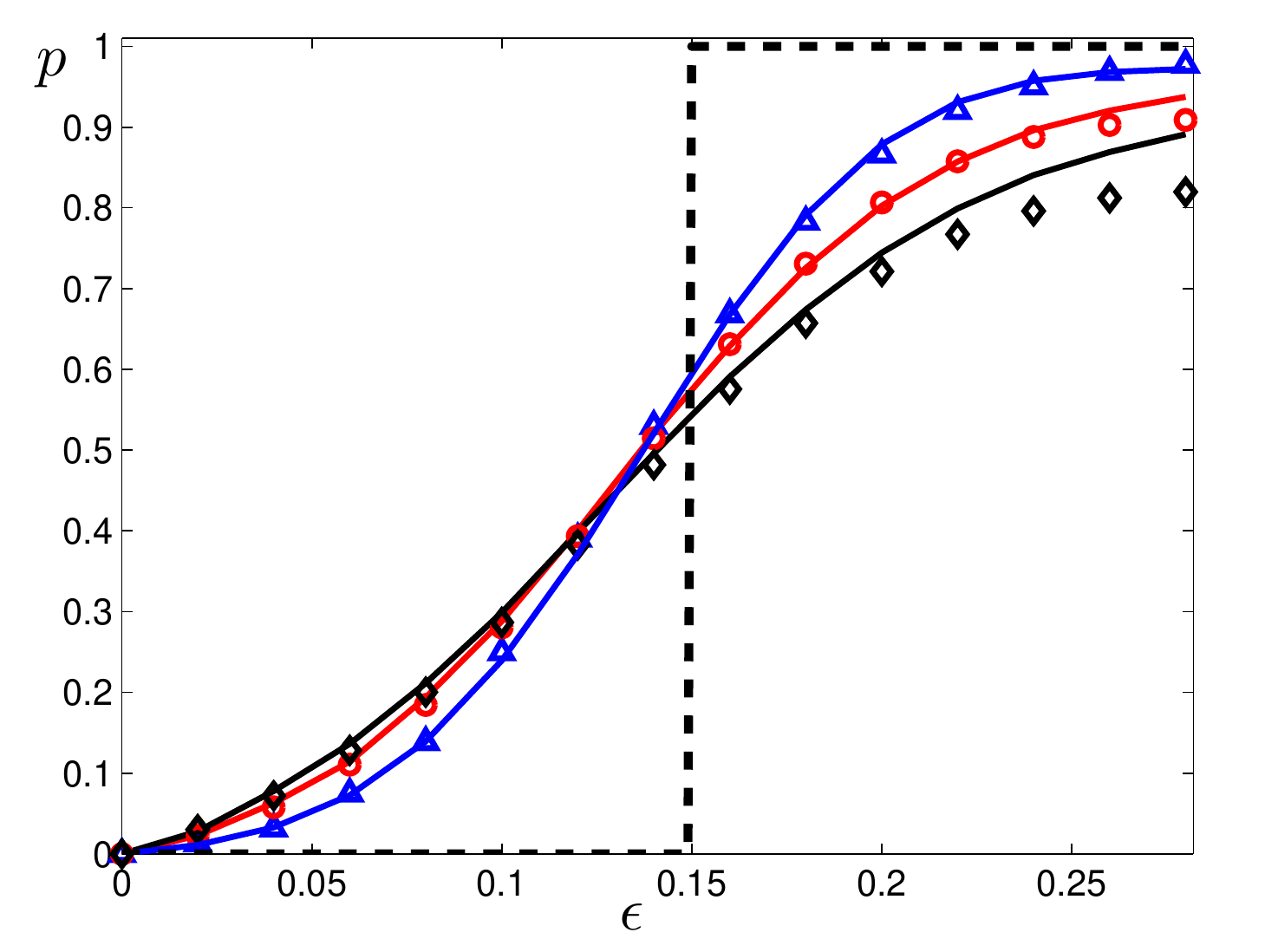}
	\caption{Probability to induce a spike by time delayed feedback for different delay force amplitudes. 
	The blue triangles, red circles, and black diamonds represent the simuation results of 
	Eq.(\ref{Eq:p_simulation}) for $D=0.005, 0.007, 0.009$ 
	[we used the Euler-Maruyama method with  
time step $dt=0.01$, integration interval was $5\times 10^5$, and additionally  
averaging over 200 realizations was performed]. The solid lines with the same color 
	is the corresponding numerical results of Eq.(\ref{Eq:p_numerical}). The black dashed line 
	is the deterministic solution with $\epsilon_{c}=0.15$. Parameters are chosen as $a=0.95, \tau=500$.}
	\label{fig:pvs_epsilon}
\end{figure}

As Fig.~\ref{fig:pvs_epsilon} depicts, the numerical method described fits well with the 
simulation results. We also investigated how the noise intensity influences the probability to induce a spike. To analyze the role of noise and delay, we compare the results in
presence of noise with the deterministic case, where there is a critical value of $\epsilon$ 
to induce periodic spikes. Generally speaking, for $\epsilon<\epsilon_{c}$, noise enhances the spiking by 
cooperation with the delay feedback, while for $\epsilon>\epsilon_{c}$ noise can prevent spikes otherwise induced by the delay feedback.
  
\subsection{Analytic approaches to calculate induced probability}

As we have shown above, the problem reduces to the analysis of a pulse-driven Fokker-Planck equation.
Such an analysis can be performed analytically in the limiting
cases of an adiabatic (very long) pulse, and of a kicked ($\delta$-function) driving. The adiabatic 
approximation appears to be rather bad,
while for a narrow pulse, as we show below, the approximation of a $\delta$-kick 
appears to be satisfactory.

It is convenient to introduce a parameter to control the width of the forcing pulse.  Therefore, 
Eq.~(\ref{Eq:model}) is modified into the following one: 
\begin{equation}
\dot{\theta}=a+\cos\theta+\epsilon C_{q} (a+\cos\theta(t-\tau))^{q}+\sqrt{D}\xi(t)\;.
\label{eq:kf}
\end{equation}
Here parameter $q$ determines the effective width of the pulse, 
and $C_{q}$ is the normalization coefficient defined as 

\[
C_{q}=\frac{1}{\int_{-\infty}^{\infty}(a+\cos\Theta_{sp}(t))^{q}dt},
\]
being consistent with Eq.~(\ref{Eq:model}) when $q=1$. 
For large values of $q$, the force in~\eqref{eq:kf} is nearly  a $\delta$-pulse. 

The analysis can be performed  in terms of the so-called splitting probability.
We start with an equilibrium solution of the autonomous Fokker-Planck 
equation~\eqref{Eq:stationary_withoutdelay}, which for small noise 
is concentrated around the stable state (minimum of the potential).
During the $\delta$ kick, the static potential and diffusion term don't play a role, and 
hence the effective evolution of
the probability density  from $\tau^{-}$ to $\tau^{+}$ is just the shift 
\begin{equation}
P(\theta,\tau^{+})=e^{-\epsilon\frac{\partial}{\partial\theta}}P(\theta,\tau^{-})
=P_{st}(\theta-\epsilon).
\end{equation}

Due to the noisy environment, the following evolution is a relaxation, described
by the autonomous Fokker-Planck equation. During this evolution, a ``particle''
can overcome the potential barrier, thus producing a spike, or return back to the 
stable state, this corresponds to not inducing a spike.
The main contribution is from the points around $\theta_s+\epsilon$, for which
we can  approximate the potential by the inverted parabolic one.  Evolution in such a potential
is known as the splitting problem \cite{gardiner2009stochastic}. If the 'phase particle' is initially 
at the position $\theta$, the probability to eventually be  right to the maximum $\theta_{u}$ is
\begin{equation}
\rho(\theta)=\frac{1}{2}\left(1-\text{erf}\left[(\theta_u-\theta)\sqrt{\frac{|U''(\theta_{u})|}{D}}\right] 
\right).
\end{equation}
Thus, the probability to induce a spike is
\begin{equation}\label{Eq:p_analytical}
p(\epsilon)=\int_{\epsilon}^{2\pi+\epsilon}P_{st}(\theta-\epsilon)\rho(\theta)d\theta \ =
\int_{0}^{2\pi}P_{st}(\theta)\rho(\theta+\epsilon)d\theta.
\end{equation}

\begin{figure}
	\includegraphics[width=0.5\textwidth]{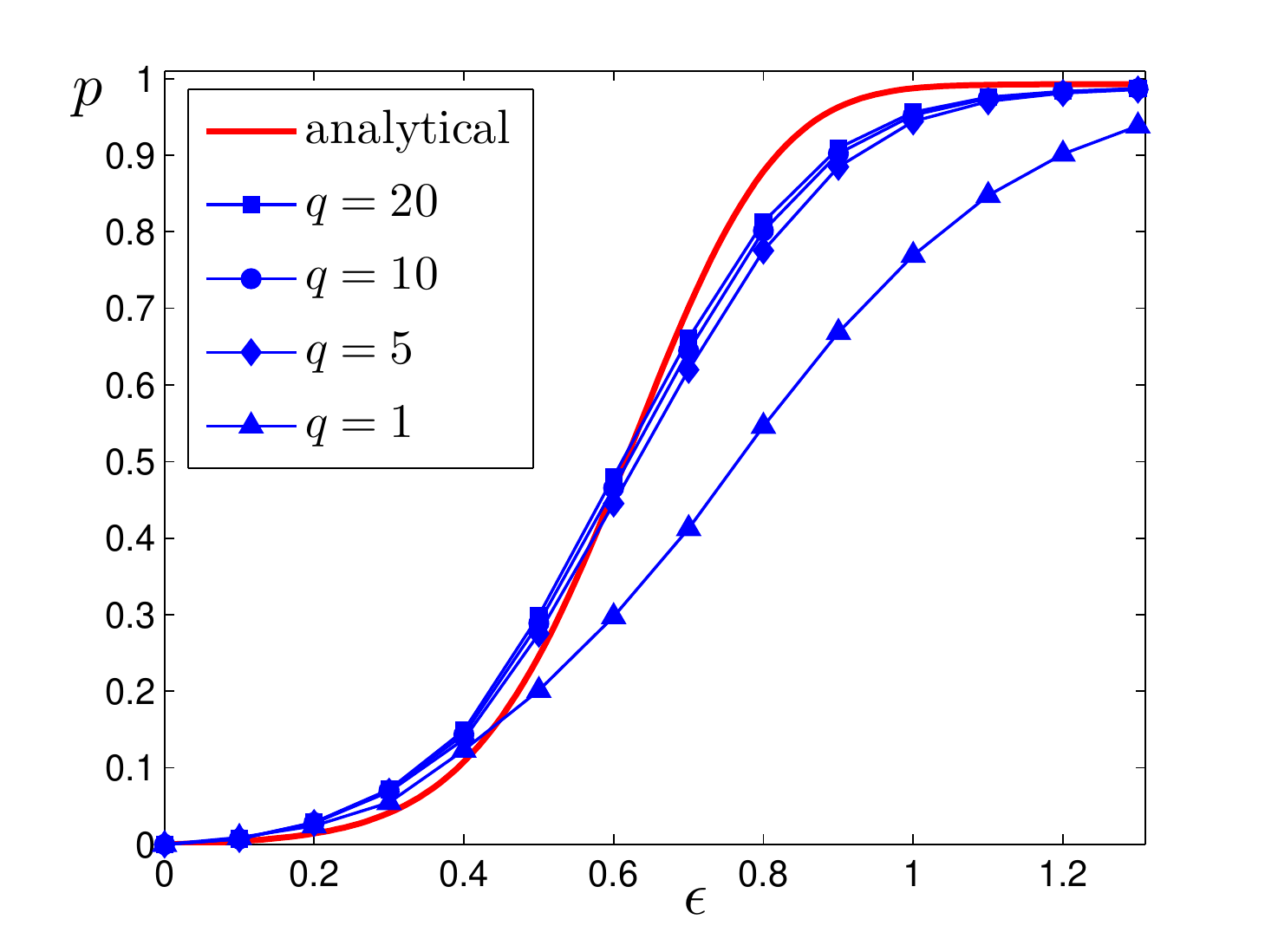}
	\caption{Probability to induce a spike by delayed pulses with different sharpness vs the amplitude 
	of the delay force. The blue triangles, diamonds, circles and rectangles represent 
	the simulation results of Eq.~(\ref{Eq:p_simulation}) with $q=1$, 5, 10 and 20 respectively.
	The red curve is the analytical result from Eq.~(\ref{Eq:p_analytical}) for 
	the $\delta$-pulse. Parameters are chosen as $a=0.995$, $D=0.005$ and $\tau=500$.}
	\label{fig:pvs_ep_nar}
\end{figure}

In  Fig.~\ref{fig:pvs_ep_nar} we compare the analytical expression for the delta-pulse with 
simulations for different values of parameter $q$.  For $q=1$ the analytic formula is not a 
good approximation, but for  $q=5$ and larger value, it fits numerics rather well.  

\section{Conclusions}
\label{sec:concl}
We have demonstrated that the combinational effect of time delay and noise can 
lead to interesting spike patterns in 
excitable neurons. We have shown that a weak 
positive (excitatory) time-delay feedback on the excitable neuron 
in a noisy environment leads to delay-induced stochastic bursting. 
As an ideal mathematical
model  to describe the spiking parttern we adopted a  
point process with the leader-follower relationship. 
The main restriction in the applicability of this model is a 
separation of time scales,
which requires noise to be weak and the delay to be long. 
The model contains just two parameters, the rate $\lambda$
of appearance of spontaneous spikes, and the probability $p$ to induce
a follower spike. Roughly, the bursting pattern can be described
as a sequence with randomly appearing busrsts (with average inter-burst interval $\lambda^{-1}$),
having random durations (as an average, each burst has $p(1-p)^{-1}$ spikes).

It is instructive to analyse the roles noise and time delay play in the model. 
When the amplitude of the delay force is below the critical value of 
onset of delay-induced
oscillations (i.e., $\epsilon<\epsilon_{c}$), noise and delay 
jointly induce spikes: delayed feedback reduces temporary the potential barrier
to overcome due to noisy forcing. On the other hand, if the amplitude of 
the delay force is above the 
critical value, i.e., $\epsilon>\epsilon_{c}$, and delay feedback is large 
enough to induce spikes in the 
deterministic case, noise makes the probability to induce spikes to be less than one, 
so that the bursts remain finite.  As a very rough estimation, one can say that exactly
at $\epsilon=\epsilon_c$ the delayed force brings the system to the unstable state (maximum
of the effective potential), from which noise can produce a spike with probability $1/2$.
This estimate is confirmed by numerical results presented in Fig.~\ref{fig:pvs_epsilon},
where the dashed line crosses the probability $p$ curves at $p\approx 1/2$.

As we have shown in the paper, two essential parameters determine statistical properties of
the stochastic bursting: the spontaneous excitation rate $\lambda$ and the probability to 
induce a spike during the feedback $p$. While the former is the standard quantity, 
easily calculated 
from the stationary solution of the autonomous Fokker-Planck 
equation, the latter probability
could be found only numerically (from the solution of forced Fokker-Planck equation)
or with some additional approximations. We have found that adiabatic approximation is
not adequate for the theta-neuron considered, while the approximation 
of a narrow, $\delta$-function-like
pulse gives a qualitatively good result. A quantitative correspondence 
could be achieved, however,
only when we modified the form of the delayed force making it narrower 
than in the original
formulation.

Our basic system in this paper was a one-dimensional 
equation similar to that of a theta-neuron. This significantly simplified the analysis
based on the Fokker-Planck equation. However, we expect that the point process
model of the dynamics will be valid in other, more realistic systems of Hodkin-Huxley type, like the
the noisy FitzHugh-Nagumo system with delayed feedback, provided the above mentioned separation
of the characteristic time scales is valid.

Finally, we hope that the approach based on the point process model can be extended to
networks of  delay-coupled noisy theta-neurons, which is one of the future subjects.

\begin{acknowledgments}
C. Z. acknowledges the financial support from China Scholarship Council (CSC). We thank Ralf 
Toenjes, Denis Goldobin and Lutz Schimansky-Geier for valuable discussions. A.P. thanks
Russian Science Foundation for support (Grant No. 17-12-01534, studies of the driven FPE).
\end{acknowledgments}


\begin{thebibliography}{18}%
\makeatletter
\providecommand \@ifxundefined [1]{%
 \@ifx{#1\undefined}
}%
\providecommand \@ifnum [1]{%
 \ifnum #1\expandafter \@firstoftwo
 \else \expandafter \@secondoftwo
 \fi
}%
\providecommand \@ifx [1]{%
 \ifx #1\expandafter \@firstoftwo
 \else \expandafter \@secondoftwo
 \fi
}%
\providecommand \natexlab [1]{#1}%
\providecommand \enquote  [1]{``#1''}%
\providecommand \bibnamefont  [1]{#1}%
\providecommand \bibfnamefont [1]{#1}%
\providecommand \citenamefont [1]{#1}%
\providecommand \href@noop [0]{\@secondoftwo}%
\providecommand \href [0]{\begingroup \@sanitize@url \@href}%
\providecommand \@href[1]{\@@startlink{#1}\@@href}%
\providecommand \@@href[1]{\endgroup#1\@@endlink}%
\providecommand \@sanitize@url [0]{\catcode `\\12\catcode `\$12\catcode
  `\&12\catcode `\#12\catcode `\^12\catcode `\_12\catcode `\%12\relax}%
\providecommand \@@startlink[1]{}%
\providecommand \@@endlink[0]{}%
\providecommand \url  [0]{\begingroup\@sanitize@url \@url }%
\providecommand \@url [1]{\endgroup\@href {#1}{\urlprefix }}%
\providecommand \urlprefix  [0]{URL }%
\providecommand \Eprint [0]{\href }%
\providecommand \doibase [0]{http://dx.doi.org/}%
\providecommand \selectlanguage [0]{\@gobble}%
\providecommand \bibinfo  [0]{\@secondoftwo}%
\providecommand \bibfield  [0]{\@secondoftwo}%
\providecommand \translation [1]{[#1]}%
\providecommand \BibitemOpen [0]{}%
\providecommand \bibitemStop [0]{}%
\providecommand \bibitemNoStop [0]{.\EOS\space}%
\providecommand \EOS [0]{\spacefactor3000\relax}%
\providecommand \BibitemShut  [1]{\csname bibitem#1\endcsname}%
\let\auto@bib@innerbib\@empty
\bibitem [{\citenamefont {Abbott}\ \emph {et~al.}(2008)\citenamefont {Abbott},
  \citenamefont {McDonnell}, \citenamefont {Pearce},\ and\ \citenamefont
  {Stocks}}]{McDonnel_etal-08}%
  \BibitemOpen
  \bibinfo {editor} {\bibfnamefont {D.}~\bibnamefont {Abbott}}, \bibinfo
  {editor} {\bibfnamefont {M.~D.}\ \bibnamefont {McDonnell}}, \bibinfo {editor}
  {\bibfnamefont {C.~E.~M.}\ \bibnamefont {Pearce}}, \ and\ \bibinfo {editor}
  {\bibfnamefont {N.~G.}\ \bibnamefont {Stocks}},\ eds.,\ \href@noop {} {\emph
  {\bibinfo {title} {Stochastic Resonance}}}\ (\bibinfo  {publisher} {CUP},\
  \bibinfo {address} {Cambridge},\ \bibinfo {year} {2008})\BibitemShut
  {NoStop}%
\bibitem [{\citenamefont {Pikovsky}\ and\ \citenamefont
	{Kurths}(1997)}]{pikovsky1997coherence}%
\BibitemOpen
\bibfield  {author} {\bibinfo {author} {\bibfnamefont {A.~S.}\ \bibnamefont
		{Pikovsky}}\ and\ \bibinfo {author} {\bibfnamefont {J.}~\bibnamefont
		{Kurths}},\ }\href@noop {} {\bibfield  {journal} {\bibinfo  {journal}
		{Physical Review Letters}\ }\textbf {\bibinfo {volume} {78}},\ \bibinfo
	{pages} {775} (\bibinfo {year} {1997})}\BibitemShut {NoStop}%
\bibitem [{\citenamefont {Goldobin}\ \emph
  {et~al.}(2003{\natexlab{a}})\citenamefont {Goldobin}, \citenamefont
  {Rosenblum},\ and\ \citenamefont
  {Pikovsky}}]{Goldobin-Rosenblum-Pikovsky-03}%
  \BibitemOpen
  \bibfield  {author} {\bibinfo {author} {\bibfnamefont {D.}~\bibnamefont
  {Goldobin}}, \bibinfo {author} {\bibfnamefont {M.}~\bibnamefont {Rosenblum}},
  \ and\ \bibinfo {author} {\bibfnamefont {A.}~\bibnamefont {Pikovsky}},\
  }\href@noop {} {\bibfield  {journal} {\bibinfo  {journal} {Phys. Rev. E}\
  }\textbf {\bibinfo {volume} {67}},\ \bibinfo {pages} {061119} (\bibinfo
  {year} {2003}{\natexlab{a}})}\BibitemShut {NoStop}%
\bibitem [{\citenamefont {Goldobin}\ \emph
  {et~al.}(2003{\natexlab{b}})\citenamefont {Goldobin}, \citenamefont
  {Rosenblum},\ and\ \citenamefont
  {Pikovsky}}]{Goldobin-Rosenblum-Pikovsky-03a}%
  \BibitemOpen
  \bibfield  {author} {\bibinfo {author} {\bibfnamefont {D.}~\bibnamefont
  {Goldobin}}, \bibinfo {author} {\bibfnamefont {M.}~\bibnamefont {Rosenblum}},
  \ and\ \bibinfo {author} {\bibfnamefont {A.}~\bibnamefont {Pikovsky}},\
  }\href@noop {} {\bibfield  {journal} {\bibinfo  {journal} {Physica A}\
  }\textbf {\bibinfo {volume} {327}},\ \bibinfo {pages} {124} (\bibinfo {year}
  {2003}{\natexlab{b}})}\BibitemShut {NoStop}%
\bibitem [{\citenamefont {Janson}\ \emph {et~al.}(2004)\citenamefont {Janson},
  \citenamefont {Balanov},\ and\ \citenamefont
  {Sch{\"o}ll}}]{janson2004delayed}%
  \BibitemOpen
  \bibfield  {author} {\bibinfo {author} {\bibfnamefont {N.~B.}\ \bibnamefont
  {Janson}}, \bibinfo {author} {\bibfnamefont {A.~G.}\ \bibnamefont {Balanov}},
  \ and\ \bibinfo {author} {\bibfnamefont {E.}~\bibnamefont {Sch{\"o}ll}},\
  }\href@noop {} {\bibfield  {journal} {\bibinfo  {journal} {Physical review
  letters}\ }\textbf {\bibinfo {volume} {93}},\ \bibinfo {pages} {010601}
  (\bibinfo {year} {2004})}\BibitemShut {NoStop}%
\bibitem [{\citenamefont {Prager}\ \emph {et~al.}(2007)\citenamefont {Prager},
  \citenamefont {Lerch}, \citenamefont {Schimansky-Geier},\ and\ \citenamefont
  {Schoell}}]{Prager_etal-07}%
  \BibitemOpen
  \bibfield  {author} {\bibinfo {author} {\bibfnamefont {T.}~\bibnamefont
  {Prager}}, \bibinfo {author} {\bibfnamefont {H.-P.}\ \bibnamefont {Lerch}},
  \bibinfo {author} {\bibfnamefont {L.}~\bibnamefont {Schimansky-Geier}}, \
  and\ \bibinfo {author} {\bibfnamefont {E.}~\bibnamefont {Schoell}},\
  }\href@noop {} {\bibfield  {journal} {\bibinfo  {journal} {J. Phys. A: Math.
  Theor.}\ }\textbf {\bibinfo {volume} {40}},\ \bibinfo {pages} {11045}
  (\bibinfo {year} {2007})}\BibitemShut {NoStop}%
\bibitem [{\citenamefont {Kouvaris}\ \emph {et~al.}(2010)\citenamefont
  {Kouvaris}, \citenamefont {Muller},\ and\ \citenamefont
  {Schimansky-Geier}}]{Kouvaris_etl-10}%
  \BibitemOpen
  \bibfield  {author} {\bibinfo {author} {\bibfnamefont {N.}~\bibnamefont
  {Kouvaris}}, \bibinfo {author} {\bibfnamefont {F.}~\bibnamefont {Muller}}, \
  and\ \bibinfo {author} {\bibfnamefont {L.}~\bibnamefont {Schimansky-Geier}},\
  }\href@noop {} {\bibfield  {journal} {\bibinfo  {journal} {Phys. Rev. E}\
  }\textbf {\bibinfo {volume} {82}},\ \bibinfo {pages} {061124} (\bibinfo
  {year} {2010})}\BibitemShut {NoStop}%
\bibitem [{\citenamefont {Goychuk}\ and\ \citenamefont
  {Goychuk}(2015)}]{Goychuk-Goychuk-15}%
  \BibitemOpen
  \bibfield  {author} {\bibinfo {author} {\bibfnamefont {I.}~\bibnamefont
  {Goychuk}}\ and\ \bibinfo {author} {\bibfnamefont {A.}~\bibnamefont
  {Goychuk}},\ }\href@noop {} {\bibfield  {journal} {\bibinfo  {journal} {New
  J. Phys.}\ }\textbf {\bibinfo {volume} {17}},\ \bibinfo {pages} {045029}
  (\bibinfo {year} {2015})}\BibitemShut {NoStop}%
\bibitem [{\citenamefont {Tsimring}\ and\ \citenamefont
  {Pikovsky}(2001)}]{tsimring2001noise}%
  \BibitemOpen
  \bibfield  {author} {\bibinfo {author} {\bibfnamefont {L.S.}~\bibnamefont
  {Tsimring}}\ and\ \bibinfo {author} {\bibfnamefont {A.}~\bibnamefont
  {Pikovsky}},\ }\href@noop {} {\bibfield  {journal} {\bibinfo  {journal}
  {Physical Review Letters}\ }\textbf {\bibinfo {volume} {87}},\ \bibinfo
  {pages} {250602} (\bibinfo {year} {2001})}\BibitemShut {NoStop}%
\bibitem [{\citenamefont {Masoller}(2003)}]{masoller2003distribution}%
  \BibitemOpen
  \bibfield  {author} {\bibinfo {author} {\bibfnamefont {C.}~\bibnamefont
  {Masoller}},\ }\href@noop {} {\bibfield  {journal} {\bibinfo  {journal}
  {Physical Review Letters}\ }\textbf {\bibinfo {volume} {90}},\ \bibinfo
  {pages} {020601} (\bibinfo {year} {2003})}\BibitemShut {NoStop}%
\bibitem [{\citenamefont {Pototsky}\ and\ \citenamefont
  {Janson}(2008)}]{Pototsky-Janson-08}%
  \BibitemOpen
  \bibfield  {author} {\bibinfo {author} {\bibfnamefont {A.}~\bibnamefont
  {Pototsky}}\ and\ \bibinfo {author} {\bibfnamefont {N.}~\bibnamefont
  {Janson}},\ }\href@noop {} {\bibfield  {journal} {\bibinfo  {journal} {Phys.
  Rev. E}\ }\textbf {\bibinfo {volume} {77}},\ \bibinfo {pages} {031113}
  (\bibinfo {year} {2008})}\BibitemShut {NoStop}%
\bibitem [{\citenamefont {Ermentrout}\ and\ \citenamefont
  {Kopell}(1986)}]{ermentrout1986parabolic}%
  \BibitemOpen
  \bibfield  {author} {\bibinfo {author} {\bibfnamefont {G.~B.}\ \bibnamefont
  {Ermentrout}}\ and\ \bibinfo {author} {\bibfnamefont {N.}~\bibnamefont
  {Kopell}},\ }\href@noop {} {\bibfield  {journal} {\bibinfo  {journal} {SIAM
  Journal on Applied Mathematics}\ }\textbf {\bibinfo {volume} {46}},\ \bibinfo
  {pages} {233} (\bibinfo {year} {1986})}\BibitemShut {NoStop}%
\bibitem [{\citenamefont {Gutkin}\ and\ \citenamefont
	{Ermentrout}(1998)}]{gutkin1998dynamics}%
\BibitemOpen
\bibfield  {author} {\bibinfo {author} {\bibfnamefont {B.~S.}\ \bibnamefont
		{Gutkin}}\ and\ \bibinfo {author} {\bibfnamefont {G.~B.}\ \bibnamefont
		{Ermentrout}},\ }\href@noop {} {\bibfield  {journal} {\bibinfo  {journal}
		{Neural computation}\ }\textbf {\bibinfo {volume} {10}},\ \bibinfo {pages}
	{1047} (\bibinfo {year} {1998})}\BibitemShut {NoStop}%
\bibitem [{\citenamefont {Lindner}\ \emph {et~al.}(2003)\citenamefont
	{Lindner}, \citenamefont {Longtin},\ and\ \citenamefont
	{Bulsara}}]{lindner2003analytic}%
\BibitemOpen
\bibfield  {author} {\bibinfo {author} {\bibfnamefont {B.}~\bibnamefont
		{Lindner}}, \bibinfo {author} {\bibfnamefont {A.}~\bibnamefont {Longtin}}, \
	and\ \bibinfo {author} {\bibfnamefont {A.}~\bibnamefont {Bulsara}},\
}\href@noop {} {\bibfield  {journal} {\bibinfo  {journal} {Neural
		computation}\ }\textbf {\bibinfo {volume} {15}},\ \bibinfo {pages} {1761}
(\bibinfo {year} {2003})}\BibitemShut {NoStop}
\bibitem [{\citenamefont {Kromer}\ \emph {et~al.}(2014)\citenamefont {Kromer},
	\citenamefont {Pinto}, \citenamefont {Lindner},\ and\ \citenamefont
	{Schimansky-Geier}}]{kromer2014noise}%
\BibitemOpen
\bibfield  {author} {\bibinfo {author} {\bibfnamefont {J.~A.}\ \bibnamefont
		{Kromer}}, \bibinfo {author} {\bibfnamefont {R.~D.}\ \bibnamefont {Pinto}},
	\bibinfo {author} {\bibfnamefont {B.}~\bibnamefont {Lindner}}, \ and\
	\bibinfo {author} {\bibfnamefont {L.}~\bibnamefont {Schimansky-Geier}},\
}\href@noop {} {\bibfield  {journal} {\bibinfo  {journal} {EPL (Europhysics
		Letters)}\ }\textbf {\bibinfo {volume} {108}},\ \bibinfo {pages} {20007}
(\bibinfo {year} {2014})}\BibitemShut {NoStop}%
\bibitem [{\citenamefont {Izhikevich}(2007)}]{izhikevich2007dynamical}%
\BibitemOpen
\bibfield  {author} {\bibinfo {author} {\bibfnamefont {E.~M.}\ \bibnamefont
		{Izhikevich}},\ }\href@noop {} {\emph {\bibinfo {title} {Dynamical systems in
			neuroscience}}}\ (\bibinfo  {publisher} {MIT press},\ \bibinfo {year}
{2007})\BibitemShut {NoStop}%
\bibitem [{\citenamefont {Luke}\ \emph {et~al.}(2014)\citenamefont {Luke},
	\citenamefont {Barreto},\ and\ \citenamefont {So}}]{luke2014macroscopic}%
\BibitemOpen
\bibfield  {author} {\bibinfo {author} {\bibfnamefont {T.~B.}\ \bibnamefont
		{Luke}}, \bibinfo {author} {\bibfnamefont {E.}~\bibnamefont {Barreto}}, \
	and\ \bibinfo {author} {\bibfnamefont {P.}~\bibnamefont {So}},\ }\href@noop
{} {\bibfield  {journal} {\bibinfo  {journal} {Frontiers in computational
			neuroscience}\ }\textbf {\bibinfo {volume} {8}},\ \bibinfo {pages} {145}
	(\bibinfo {year} {2014})}\BibitemShut {NoStop}%
\bibitem [{\citenamefont {Park}\ and\ \citenamefont {Kim}(1996)}]{Park-Kim-96}%
\BibitemOpen
\bibfield  {author} {\bibinfo {author} {\bibfnamefont {S.~H.}\ \bibnamefont
		{Park}}\ and\ \bibinfo {author} {\bibfnamefont {S.}~\bibnamefont {Kim}},\
}\href {\doibase 10.1103/PhysRevE.53.3425} {\bibfield  {journal} {\bibinfo
	{journal} {Phys. Rev. E}\ }\textbf {\bibinfo {volume} {53}},\ \bibinfo
{pages} {3425} (\bibinfo {year} {1996})}\BibitemShut {NoStop}%
\bibitem [{\citenamefont {Tessone}\ \emph {et~al.}(2007)\citenamefont
	{Tessone}, \citenamefont {Scir\`e}, \citenamefont {Toral},\ and\
	\citenamefont {Colet}}]{Tessone_etal-07}%
\BibitemOpen
\bibfield  {author} {\bibinfo {author} {\bibfnamefont {C.~J.}\ \bibnamefont
		{Tessone}}, \bibinfo {author} {\bibfnamefont {A.}~\bibnamefont {Scir\`e}},
	\bibinfo {author} {\bibfnamefont {R.}~\bibnamefont {Toral}}, \ and\ \bibinfo
	{author} {\bibfnamefont {P.}~\bibnamefont {Colet}},\ }\href {\doibase
	10.1103/PhysRevE.75.016203} {\bibfield  {journal} {\bibinfo  {journal} {Phys. Rev. E}\ }\textbf {\bibinfo {volume} {75}},\ \bibinfo {pages} {016203}
	(\bibinfo {year} {2007})}\BibitemShut {NoStop}%
\bibitem [{\citenamefont {Zaks}\ \emph {et~al.}(2003)\citenamefont {Zaks},
	\citenamefont {Neiman}, \citenamefont {Feistel},\ and\ \citenamefont
	{Schimansky-Geier}}]{Zaks_etal-03}%
\BibitemOpen
\bibfield  {author} {\bibinfo {author} {\bibfnamefont {M.~A.}\ \bibnamefont
		{Zaks}}, \bibinfo {author} {\bibfnamefont {A.~B.}\ \bibnamefont {Neiman}},
	\bibinfo {author} {\bibfnamefont {S.}~\bibnamefont {Feistel}}, \ and\
	\bibinfo {author} {\bibfnamefont {L.}~\bibnamefont {Schimansky-Geier}},\
}\href {\doibase 10.1103/PhysRevE.68.066206} {\bibfield  {journal} {\bibinfo
	{journal} {Phys. Rev. E}\ }\textbf {\bibinfo {volume} {68}},\ \bibinfo
{pages} {066206} (\bibinfo {year} {2003})}\BibitemShut {NoStop}%
\bibitem [{\citenamefont {Sonnenschein}\ \emph {et~al.}(2013)\citenamefont
	{Sonnenschein}, \citenamefont {Zaks}, \citenamefont {Neiman},\ and\
	\citenamefont {Schimansky-Geier}}]{Sonnenschein-13}%
\BibitemOpen
\bibfield  {author} {\bibinfo {author} {\bibfnamefont {B.}~\bibnamefont
		{Sonnenschein}}, \bibinfo {author} {\bibfnamefont {M.}~\bibnamefont {Zaks}},
	\bibinfo {author} {\bibfnamefont {A.}~\bibnamefont {Neiman}}, \ and\ \bibinfo
	{author} {\bibfnamefont {L.}~\bibnamefont {Schimansky-Geier}},\ }\href
{\doibase 10.1140/epjst/e2013-02034-7} {\bibfield  {journal} {\bibinfo
		{journal} {The European Physical Journal Special Topics}\ }\textbf {\bibinfo
		{volume} {222}},\ \bibinfo {pages} {2517} (\bibinfo {year}
	{2013})}\BibitemShut {NoStop}%
\bibitem [{\citenamefont {Ionita}\ and\ \citenamefont
	{Meyer-Ortmanns}(2014)}]{ionita2014physical}%
\BibitemOpen
\bibfield  {author} {\bibinfo {author} {\bibfnamefont {F.}~\bibnamefont
		{Ionita}}\ and\ \bibinfo {author} {\bibfnamefont {H.}~\bibnamefont
		{Meyer-Ortmanns}},\ }\href@noop {} {\bibfield  {journal} {\bibinfo  {journal}
		{Physical review letters}\ }\textbf {\bibinfo {volume} {112}},\ \bibinfo
	{pages} {094101} (\bibinfo {year} {2014})}\BibitemShut {NoStop}%
\bibitem [{\citenamefont {B{\"o}rgers}\ and\ \citenamefont
	{Kopell}(2005)}]{Borgers-Koppell-05}%
\BibitemOpen
\bibfield  {author} {\bibinfo {author} {\bibfnamefont {C.}~\bibnamefont
		{B{\"o}rgers}}\ and\ \bibinfo {author} {\bibfnamefont {N.}~\bibnamefont
		{Kopell}},\ }\href@noop {} {\bibfield  {journal} {\bibinfo  {journal} {Neural
			Comp.}\ }\textbf {\bibinfo {volume} {17}},\ \bibinfo {pages} {557} (\bibinfo
	{year} {2005})}\BibitemShut {NoStop}%
%
\bibitem [{\citenamefont {Risken}(1996)}]{risken1996fokker}%
  \BibitemOpen
  \bibfield  {author} {\bibinfo {author} {\bibfnamefont {H.}~\bibnamefont
  {Risken}},\ }\href@noop {} {\emph {\bibinfo {title} {The Fokker-Planck
  Equation}}}\ (\bibinfo  {publisher} {Springer},\ \bibinfo {year}
  {1996})\BibitemShut {NoStop}%
\bibitem [{\citenamefont {Cox}(1967)}]{cox1967renewal}%
  \BibitemOpen
  \bibfield  {author} {\bibinfo {author} {\bibfnamefont {D.~R.}\ \bibnamefont
  {Cox}},\ }\href@noop {} {\emph {\bibinfo {title} {Renewal theory}}},\
  Vol.~\bibinfo {volume} {1}\ (\bibinfo  {publisher} {Methuen London},\
  \bibinfo {year} {1967})\BibitemShut {NoStop}%
\bibitem [{\citenamefont {Gerstner}\ \emph {et~al.}(2014)\citenamefont
  {Gerstner}, \citenamefont {Kistler}, \citenamefont {Naud},\ and\
  \citenamefont {Paninski}}]{gerstner2014neuronal}%
  \BibitemOpen
  \bibfield  {author} {\bibinfo {author} {\bibfnamefont {W.}~\bibnamefont
  {Gerstner}}, \bibinfo {author} {\bibfnamefont {W.~M.}\ \bibnamefont
  {Kistler}}, \bibinfo {author} {\bibfnamefont {R.}~\bibnamefont {Naud}}, \
  and\ \bibinfo {author} {\bibfnamefont {L.}~\bibnamefont {Paninski}},\
  }\href@noop {} {\emph {\bibinfo {title} {Neuronal dynamics: From single
  neurons to networks and models of cognition}}}\ (\bibinfo  {publisher}
  {Cambridge University Press},\ \bibinfo {year} {2014})\BibitemShut {NoStop}%
\bibitem [{\citenamefont {Stratonovich}(1967)}]{stratonovich1967topics}%
  \BibitemOpen
  \bibfield  {author} {\bibinfo {author} {\bibfnamefont {R.~L.}\ \bibnamefont
  {Stratonovich}},\ }\href@noop {} {\emph {\bibinfo {title} {Topics in the
  theory of random noise}}}\ (\bibinfo  {publisher} {CRC Press},\ \bibinfo
  {year} {1967})\BibitemShut {NoStop}%
\bibitem [{\citenamefont {Gardiner}(2009)}]{gardiner2009stochastic}%
  \BibitemOpen
  \bibfield  {author} {\bibinfo {author} {\bibfnamefont {C.}~\bibnamefont
  {Gardiner}},\ }\href@noop {} {\emph {\bibinfo {title} {Stochastic methods}}}\
  (\bibinfo  {publisher} {springer Berlin},\ \bibinfo {year}
  {2009})\BibitemShut {NoStop}%
\end{thebibliography}
\end{document}